# Impact of misfit strain on the properties of tetragonal Pb(Zr,Ti)O$_3$ thin film heterostructures

Ludwig Geske,[a)] I. B. Misirlioglu,[b)] Ionela Vrejoiu, Marin Alexe, and Dietrich Hesse
*Max Planck Institute of Microstructure Physics, Weinberg 2, D-06120 Halle, Germany*

**Abstract**

Heterostructures consisting of PbZr$_{0.2}$Ti$_{0.8}$O$_3$ and PbZr$_{0.4}$Ti$_{0.6}$O$_3$ films grown on a SrTiO$_3$ (100) substrate with a SrRuO$_3$ bottom electrode were prepared by pulsed laser deposition. Using the additional interface provided by the ferroelectric bilayer structure and changing the sequence of the layers, the dislocation content and domain patterns were varied. The resulting microstructure was investigated by transmission electron microscopy. Macroscopic ferroelectric measurements have shown a large impact of the formation of dislocations and 90° domains on the ferroelectric polarization and dielectric constant. A thermodynamic analysis using the LANDAU-GINZBURG-DEVONSHIRE approach that takes into account the ratio of the thicknesses of the two ferroelectric layers and electrostatic coupling is used to describe the experimental data.

---

[a)] Electronic mail: lgeske@mpi-halle.de
[b)] Present Adress: Nuclear Science&Engineering, Massachusetts Institute of Technology, Northwest Bldg. 13, Cambridge, MA, USA

## Introduction

Ferroelectric thin films offer a variety of possible applications,[1-4] including capacitors, pyroelectric sensors, FeRAMs and valves for ink, fuel or medicines. In order to integrate ferroelectrics into suitable devices miniaturization is frequently necessary. At some critical size strains occurring at interfaces become important,[5] enabling strain engineering of the ferroelectric properties,[6] e.g. by growing on different substrates.[7] Depending on the used substrate-film combination, either compressive or tensile strains can be introduced, the latter being able to tilt the polarization vector from the out-of-plane into the in-plane direction.[8] Furthermore the polarization can be increased via strain-polarization coupling,[9] even though this is not always as extensive as expected from the increased tetragonality.[10-13] In this way ferroelectric films can be tuned to exhibit either polarization values superior to the corresponding bulk material or an outstanding dielectric constant. Other properties like the pyroelectric effect are affected as well.[14-16]

However these considerations only hold true for a very confined thickness range. If a critical thickness is exceeded during film growth, the film starts to relax by forming misfit and threading dislocations.[17-22] Additional stress arises when cooling down the film from growth temperature to room temperature due to different thermal expansion coefficients between film and underneath substrate. For certain compositions, $a$-domains can form below the CURIE temperature to further relax the residual stresses.[23] Any undesirable strains evolving from interfaces and from dislocations can be detrimental for the ferroelectric behavior.[24]

Another approach to tune the properties is to grow bilayers or superlattices which combine ferroelectrics with other classes of material, e.g. semi-[25] or superconductors.[26] By combining systems with very similar crystallographic properties like ferroelectric $PbTiO_3$ and paraelectric $SrTiO_3$ (*STO*) intriguing effects such as very high dielectric constants for a critical thickness ratio are predicted.[27] On the other hand, the presence of such a high dielectric anomaly due to the transition of the ferroelectric layer to the paraelectric phase at a critical fraction of the paraelectric layer is now under debate. Some recent studies [28-30] demonstrate that this critical fraction can be perceived as the point at which the ferroelectric layer can no longer exist in the single domain state but it will split into 180° electrical domains, equivalent to a thermodynamically more stable phase. Therefore an intrinsic dielectric anomaly will not be exhibited.

In this study, bilayer heterostructures consisting of two tetragonal $Pb(Zr,Ti)O_3$ (*PZT*) compositions $PbZr_{0.2}Ti_{0.8}O_3$ (*PZT20/80*) and $PbZr_{0.4}Ti_{0.6}O_3$ (*PZT40/60*) are discussed. The influence of the interface between the ferroelectric layers on the resulting macroscopic electric properties, together with the resulting strains, dislocation states and domains are investigated. Experimental film growth, microstructural and electrical characterization are followed by a LANDAU-GINZBURG-DEVONSHIRE (LGD) approach to interpret the results and to shed light on the impact of $a$-domains on such bilayer structures. It is shown, that $a$-domains in bilayers and superlattices can arise under certain strain conditions and can significantly alter the electrical properties. The strain states in the layers can be adjusted by changing the sequence of layer growth or by choosing particular thickness ratios of the layers.

## Experimental

Pulsed laser deposition (PLD) was used to grow thin film heterostructures on vicinal (100) *STO* single crystals with a miscut of about 0.1° (CrysTec, Berlin/Germany). $TiO_2$-terminated surfaces with atomically smooth terraces were obtained by etching the STO substrate in buffered hydrofluoric acid [31] and subsequently annealing at 1100 °C for one hour.[32] The ferroelectric *PZT20/80 / PZT40/60* bilayers were successively grown on top of the $SrRuO_3$ (*SRO*) bottom electrode, which was grown first on STO (100) in step-flow growth mode,[33] using a substrate temperature range of 575-700°C, an oxygen pressure of 14-30 Pa, a laser fluence of 2.5-5 J/cm² and a repetition rate of 5 Hz. Circular Pt top electrodes with a diameter of about

100 μm were deposited at room temperature by RF sputtering through a corresponding stencil. Macroscopic characterization comprised ferroelectric hysteresis curves recorded at 1 kHz (AixxACT TF Analyzer) and capacitance-voltage characteristics measured at 100 kHz with a probing voltage of 0.1 V (HP4194A Impedance Analyzer). Structure analysis was performed by transmission electron microscopy on cross-section samples employing a Philips CM20T electron microscope at 200 keV primary electron energy, using the *STO* [010] direction as the one of the incident beam.

### Results

At a given film-substrate lattice misfit the dislocation content and domain formation in single composition thin films are influenced mainly by the film thickness and the growth conditions. A bilayer structure offers the possibility to affect both features by the presence of the additional interface. Due to the different misfits between the layers and between the individual layers and the substrate various relaxation and domain states are possible.

In this study, a system containing a *STO* substrate together with mainly c-axis oriented *PZT20/80* and *PZT40/60* layers was chosen because of their relative small lattice misfit. At room temperature, *PZT20/80* and *PZT40/*60 have a pseudocubic misfit with the *STO* substrate of f = -2% and f = -3.2%, respectively. In all present experiments a *SRO* film was used as bottom electrode. *SRO* has a misfit of f = -0.5% and grows pseudomorphically to the *STO* when thinner than ca. 75 nm.[22] Therefore, the misfit of the *PZT* layers can be treated as they were grown directly on *STO*.

There are mainly two possibilities shown schematically in Fig.1. (i) When the first grown layer is *PZT20/80*, this is strained to the substrate; thus the subsequent *PZT40/60* layer grows by forming misfit dislocations (MDs) at the interface accompanied by threading dislocations (TDs) propagating to the top surface. In addition, the top layer exhibits *a*-domains which are also terminated at the interface. TEM pictures depicting this case are shown in Fig.1a together with a schematic drawing in Fig.1b. (ii) When *PZT40/60* is used as the bottom layer, MDs are immediately formed at the interface with the *SRO* electrode from which many TDs propagate to the top surface of the structure, thereby crossing the *PZT20/80* top layer. If the strain state changes at the interface it also acts as a barrier for the TDs' propagation,[34,35] and reduces the dislocation content in the top layer with respect to the bottom one. Moreover, two different domain states are possible in the case of this particular dislocation distribution: 1) the *a/c*-domains are confined to the *PZT20/80* layer and terminate at the interface, as shown in Fig.1c-d; 2) the domains are crossing the interface and penetrate through the entire film (Fig.1e-f) in order to reduce the overall elastic energy of the structure, when the elastic energy of the partially strained film is high enough (possible in thicker films).

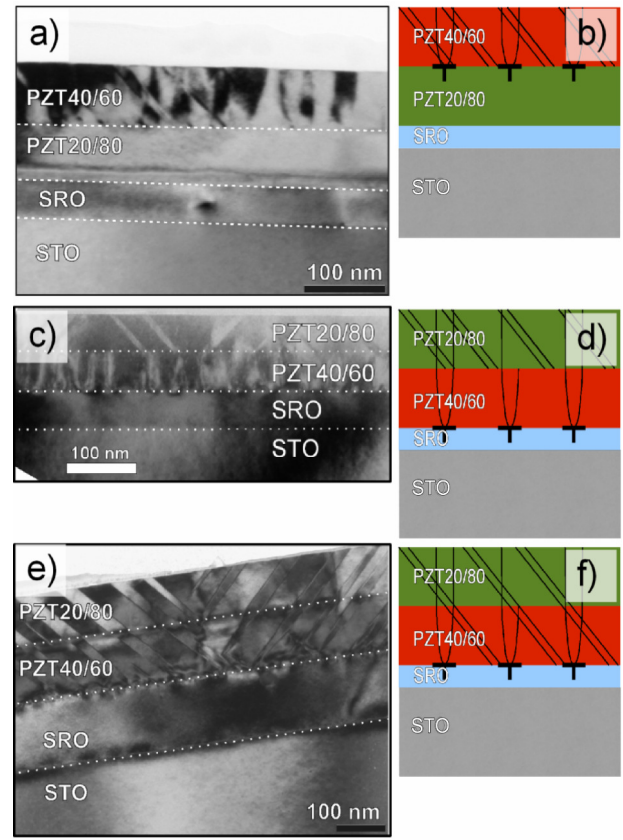

**Fig.1:** (Color online) TEM cross-section micrographs (a, c, e) and according shemes (b, d, f) of ferroelectric bilayers consisting of PZT20/80 and PZT40/60 grown with a SRO bottom electrode on (001)-oriented STO, seen from the [010] STO direction.

The dependence of the remnant polarization $P_r$ and dielectric constant $\varepsilon_r$ on the relative thickness $\alpha = t_{PZT40/60} / t_{film}$ with $t_{film} = t_{PZT40/60} + t_{PZT20/80}$, of the structures are shown in Fig.2. It can be seen that the different microstructures significantly modify the values of measured $P_r$ and $\varepsilon_r$. Structures with a *PZT20/80* bottom layer containing a negligible density of dislocations (Fig.1a-b and corresponding open circles in Fig.2) exhibit mean values of $P_r \approx 70$ µC/cm² and $\varepsilon_r \approx 145$. In contrast to this, the films with a dislocation-rich *PZT40/60* bottom layer (Fig.1c-f and corresponding full circles in Fig.2) show a smaller $P_r$ of about 35 µC/cm² and a much higher $\varepsilon_r \approx 435$. The codomains caused by the two possible sequences in the bilayers are indicated by the shaded areas in Fig.2.

The LGD theory for ferroelectrics was employed in an attempt to understand the observed experimental results. It included appropriate modifications taking into account the misfit strain due to the film-substrate lattice mismatch and the electrostatic coupling of the ferroelectric layers. As the layers are well above the usual thickness for similar systems where interface- and size-effect related phenomena have been reported, such effects have been neglected. The free energy density of a bilayer is described by [27]

$$F = \alpha F_1 + (1-\alpha)F_2 + F_c \quad (1)$$

with the relative thickness $\alpha$, the energies $F_i$ of the individual layers and an additional contribution $F_c$ due to the electrostatic coupling between the layers. The energy densities $F_i$ can be written in the form

$$F_i = F_0 + aP^2 + bP^4 + cP^6 - EP \quad (2)$$

where $a$, $b$ and $c$ are the thermodynamic coefficients, P the polarization and E the external electric field parallel to the polarization. $c$ is the higher order dielectric stiffness coefficient $\alpha_{111}$. $a$ and $b$ coefficients have to be modified in order to include the effect of the pseudocubic misfit and the clamping between the thin film and the substrate. For different domain states, different forms of the coefficients $a$ and $b$ are introduced in the energy density $F$.[36] If the crystal structure contains only *c*-domains,

$$a = \frac{T - T_C}{2\varepsilon_0 C} - f \frac{2Q_{12}}{S_{11} + S_{12}}; \quad b = \alpha_{11} + \frac{Q_{12}^2}{S_{11} + S_{12}} \quad (3)$$

with $T_C$ being the CURIE temperature, $C$ the CURIE constant, $S_{ij}$ the elastic compliances, $Q_{ij}$ the electrostrictive coefficients and $\alpha_{11}$ a higher order dielectric stiffness coefficient. The coefficients for a structure containing *a/c*- and $a_1/a_2$-domains are

$$a^* = \frac{T - T_C}{2\varepsilon_0 C} - f \frac{Q_{12}}{S_{11}}; \quad b^* = \alpha_{11} + \frac{Q_{12}^2}{2S_{11}} \quad (4)$$

and

$$a^{**} = \frac{T - T_C}{2\varepsilon_0 C} - f \frac{Q_{11} + Q_{12}}{S_{11} + S_{12}},$$
$$b^{**} = \alpha_{11} + \frac{(Q_{11} + Q_{12})^2}{4(S_{11} + S_{12})} \quad (5)$$

respectively. In order to decide which domain configuration is stable for a given misfit strain, the free energy has to be completed with the term describing the misfit contribution. This term is

$$\frac{f^2}{S_{11} + S_{22}} \quad (6)$$

for the *c*- and the $a_1/a_2$-domain configuration and

$$\frac{f^2}{2S_{11}} \quad (7)$$

for the *a/c*-domain configuration. The minimization of the free energy will give the stable domain configuration. An important term in the free energy of the ferroelectric heterostructures is the one describing the electrostatic coupling between the component layers. This term increases the energy due to polarization difference at the interface. For the structures shown in Fig.1a-d the model should include a single-domain bottom layer and a multi-domain top layer. If there are only *c*-domains in both layers, then the coupling term reads

$$F_C = \frac{1}{2\varepsilon_o} \alpha(1-\alpha)(P_1 - P_2)^2 \quad (8)$$

with $\varepsilon_0$ being the dielectric permittivity of vacuum, $P_1$ the polarization of the top layer

(layer 1) and $P_2$ the polarization of the bottom layer (layer 2). In case of an *a/c*-domain configuration of the top layer, the fraction $\Phi_a$ of *a*-domains will be determined by

$$\Phi_a = \frac{(S_{11} - S_{12})(f - Q_{12}P_{c1}^2)}{S_{11}(Q_{11} - Q_{12})P_{c1}^2} \quad (9).$$

The single *c*-domain state of the bottom layer induces a *c*-component of the polarization in *a*-domains of the top layer and couples to the *c*-domain as in (8). Therefore the electrostatic coupling can be described as

$$F_C = \frac{1}{2\varepsilon_0}\alpha(1-\alpha)\big((1-\phi_a)P_{c1} + \phi_a P_{a1} - P_2\big)^2. \quad (10)$$

Here $P_{c1}$ is the polarization of the *c*-domains and $P_{a1}$ is the induced *c*-polarization in the *a*-domains of layer 1. In case of the electrostatic coupling with the $a_1/a_2$-domain configuration of the top layer, the induced polarization regards the whole layer 1 and the coupling term can be written as

$$F_C = \frac{1}{2\varepsilon_0}\alpha(1-\alpha)(P_{a1} - P_2)^2. \quad (11)$$

If both layers exhibit an *a/c*-domain structure the coupling term becomes

$$F_C = \frac{1}{2\varepsilon_0}\alpha(1-\alpha)\big((1-\phi_{a1})P_{c1} + \phi_{a1}P_{a1} - (1-\phi_{a2})P_{c2} - \phi_{a2}P_{a2}\big)^2 \quad (12)$$

with $\Phi_{a1}$ and $\Phi_{a2}$ the fraction of *a*-domains in the first and the second layer, respectively. The induced *c*-polarization in the *a*-domains gives rise to an additional energy term that also has to be taken into account. This can be deduced from the general formulation given by Pertsev *et al.*[37], thus by equating both in-plane components of the polarization:

$$F_3(P, E=0) = 2a^{**}P_a^2 + aP_c^2 + b_1P_a^4 + bP_c^4 + b_2P_a^2P_c^2 + \alpha_{111}(2P_a^6 + P_c^6) + \alpha_{112}(2P_a^4(P_a^2 + P_c^2) + 2P_c^4P_a^2) + \alpha_{123}P_a^4P_c^2 \quad (13)$$

containing the higher order dielectric stiffness coefficients $\alpha_{ijk}$ and the modified coefficients

$$b_1 = \left(\alpha_{11} + \frac{1}{2}\frac{(Q_{11}^2 + Q_{12}^2)S_{11} - 2Q_{11}Q_{12}S_{12}}{S_{11}^2 - S_{12}^2}\right) + \left(\alpha_{12} - \frac{(Q_{11}^2 + Q_{12}^2)S_{12} - 2Q_{11}Q_{12}S_{11}}{S_{11}^2 - S_{12}^2} + \frac{Q_{44}^2}{2S_{44}}\right) \quad (14)$$

and

$$b_2 = \alpha_{12} + \frac{Q_{12}(Q_{11} + Q_{12})}{S_{11} + S_{12}}. \quad (15)$$

The stable equilibrium domain configuration can be determined using the relations (1)-(15). $F$ has to be minimized in order to calculate the remnant polarization $P_r$. The small signal dielectric constant $\varepsilon_r$ results from the polarization difference when applying a small external electric field $E_0$:

$$\varepsilon_r = \frac{P(E=E_0) - P(E=0)}{E_0}. \quad (16)$$

In order to compare the measured values (given by the dots in Fig.2) with the theoretical description, the strain states of the different layers must be known. Since these are quite difficult to determine experimentally and change from sample to sample, only some special cases will be considered in the calculations and will be compared with the experimental data to visualize the possible range. The first considered case is a bilayer with a fully strained *PZT40/60* layer (misfit at room temperature: $f_{RT}$ = -3.2 %) on top of a fully strained *PZT20/80* layer ($f_{RT}$ = -2.0 %). The corresponding values for polarization and dielectric constant are given in Fig.3a and 3b by the red dotted line no. 1. However, the TEM picture in Fig.1a shows a lot of TDs in the top *PZT40/60* layer suggesting a mechanical relaxation of the layer. In the extreme case this layer can be treated as fully relaxed at growth temperature ($f_{RT}$ = -0.1 %). The results are shown by line no. 2 in Fig.3a and b, where $P_r$ is smaller and $\varepsilon_r$ larger compared to line no. 1. In reality, both layers will partially relax to some point, which is determined by the PLD growth conditions which can not be completely controlled or exactly measured. It has to be assumed that the measured values lie somewhere in the range between the two calculated red dotted lines. Concerning the *PZT20/80* on *PZT40/60* bilayer with domains terminated at the interface (Fig.1c), the curves no. 3 and no. 4 (black lines) show the results of a relaxed *PZT20/80* ($f_{RT}$ = -0.1 %) on a relaxed *PZT40/60* ($f_{RT}$ = -0.1 %) and of a strained *PZT20/80* ($f_{RT}$ = +1.1 %) on a relaxed *PZT40/60* layer, respectively. In this case, the film containing a strained *PZT20/80* layer exhibits a smaller $P_r$ and a larger $\varepsilon_r$. The lines denoted as 3´ and 4´ cover the possibility of domains to propagate through both layers as shown in Fig.1e. It can be seen that the influence of the *a/c*-domain structure on $P_r$ is small while $\varepsilon_r$ increases considerably.

## Discussion

Although the properties and lattice constants of the tetragonal *PZT* compositions *PZT20/80* and *PZT40/60* are very similar, the combination of both in the form of bilayers results in very different values for the remnant polarization $P_r$ and the dielectric constant $\varepsilon_r$ when the layer sequence with respect to the substrate is changed. The main reasons for this behavior are 1) the different lattice parameters of the two *PZT* compositions and 2) the dependence of the misfit strain of the top layer on the relaxation state of the bottom layer. Concerning the growth on the *STO* (100) substrate, the lattice constant of *PZT20/80* is close enough to allow a pseudomorphic growth (for films thinner than ca. 100 nm), whereas *PZT40/60* forms dislocations to release the strain caused by its higher lattice mismatch. Furthermore the domain and polarization states of the two layers have to adjust to each other. The interface between the ferroelectric layers is the place of the mechanical and electrostatic coupling and it can, therefore, act as a barrier or nucleation site for the formation of domains and dislocations, allowing different domain states and dislocation densities in the two layers.

It has been shown that the observed trends can be reproduced by the LGD analysis. Ferroelectric bilayers containing *PZT20/80* as bottom layer, hence with both layers subjected to compressive stress, show high polarization values and a low dielectric constant (curve no. 1 in Fig.3). The consecutive relaxation of the *PZT40/60* (curve no. 2) and of the *PZT20/80* layer (curve no. 3) leads to a decrease in $P_r$ and increase in $\varepsilon_r$ due to the *a*-domains and the domain wall contribution.[38]

If a *PZT40/60* bottom layer is used, even tensile stress can occur for the *PZT20/80* layer. That would not happen if the *PZT20/80* layer grew directly on the *SRO*-coated *STO* substrate but becomes possible because of the bilayer structure. In this case, the $P_r$ would further decrease and $\varepsilon_r$ further increase (curve no. 4) compared to states with less tensile stress. As it is shown in Fig.1e, the domains might also cross the interface. This causes a slight increase of $P_r$ and a significant increase of $\varepsilon_r$ (curve no. 3' and 4') due to the further relaxation and the contributions of the *a/c*-domain structure compared with the films containing the untwinned *PZT40/60* bottom layer.

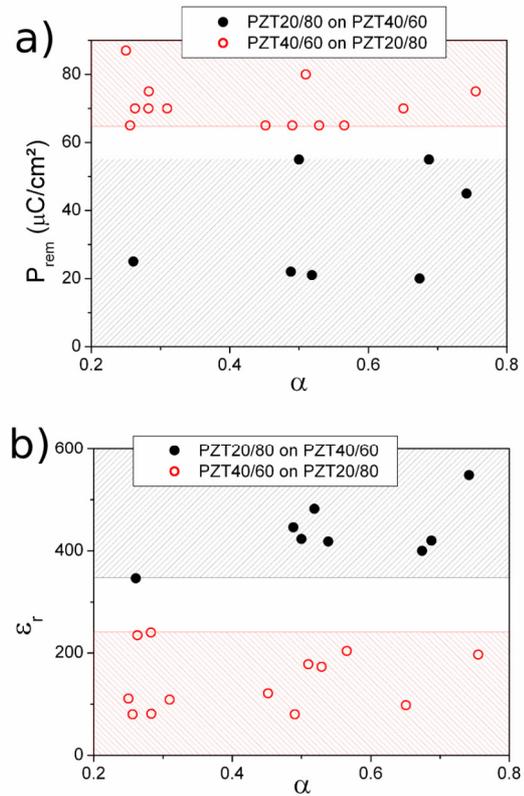

**Fig.2:** (Color online) Remnant polarization (a) and dielectric constant (b) of bilayers with a PZT20/80 (○) and a PZT40/60 bottom layer (●) in dependence on the relative thickness. The shaded areas designate the codomains of the measured values caused by the different layer sequences.

If $\alpha$ becomes zero or one, the film entirely consists either of *PZT20/80* or of *PZT40/60*, respectively. At $\alpha = 0$ the values correspond to a *PZT20/80* film under compression (curve no. 1 and 2), no stress (curve no. 3 and 3') and tension (curve no. 4 and 4'), respectively. On the other hand at $\alpha = 1$ the values for a *PZT40/60* film subjected to compressive stress (curve no. 1) and no stress with (curve no. 2, 3', 4') and without *a*-domains (curve no. 3 and 4) can be read off. These results can be compared with measurement data obtained on single layer films (■ and □ in Fig.3). It turns out that the $P_r$ value of a relaxed *PZT20/80* (designated with B) is described very well by the calculations, whereas the measured $P_r$ value of a strained *PZT20/80* layer (A) is much higher. This phenomenon has already been observed in a former work.[39] The simulated value for a *PZT40/60* layer (C) also gives slightly smaller values than the measurement. Concerning $\varepsilon_r$ there is a good agreement between simulation and experiment for *PZT20/80* and the calculated range includes the measured value for *PZT40/60*.

Despite the good agreement between the results from the LGD theory and the experiment there are still some deviations. These occur because the model used is still quite simple in comparison to the diversity of the features of the investigated system. It does not take into account local microstresses and possible internal fields originating from these microstresses. The major influences considered by the model are the misfit strain and the overall electrostatic coupling between the layers. For a complete model additional effects induced by the interface between the ferroelectric layers and by the interfaces with the metal electrodes should be taken into account, not mentioning the polarization fluctuations that should be expected to contribute to the system energy.

The depletion region which might occur at the metal-*PZT* interface was not considered.[40] Charged traps can also significantly contribute to $\varepsilon_r$.[41] The presence of the "dead layer" at the interfaces may significantly change the ferroelectric properties.[42,43] The interface between the ferroelectric layers might carry space charges, which can also change the properties of the bilayer.[44] Misfit dislocations which form at the interface are accompanied by local strains which affect both $P_r$ and $\varepsilon_r$,[24,45,46] and give rise to threading dislocations.[15] These threading dislocations in turn have been only taken into account as a relaxation mechanism, but they also directly affect $P_r$.[43,47] Overall, despite the simplicity of the approach, the variations of the experimental observations can be elucidated and the effect of *a*-domains can be highlighted through the adopted methodology.

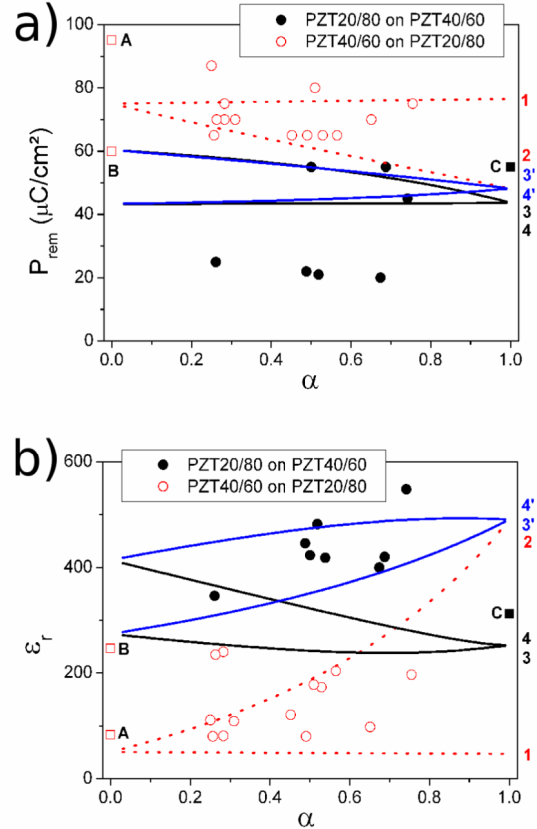

**Fig.3:** (Color online) Remnant polarization (a) and dielectric constant (b) of bilayers with a PZT20/80 (○) and a PZT40/60 bottom layer (●) in dependence on the relative thickness. □ and ■ designate single PZT layers consisting of strained PZT20/80 (A), relaxed PZT20/80 (B) and relaxed PZT40/60 (C). The lines display the results of the LGD theory for bilayers with a PZT20/80 bottom layer (red dotted line, 1, 2) and a PZT40/60 bottom layer with (blue continuous line, 3', 4') and without a/c domain walls (black continuous line, 3, 4)

## Summary

Different dislocation and domain states were induced in *PZT20/80* / *PZT40/60* bilayers grown on *SRO*-coated *STO* (100) by changing the growth sequence and the thickness of the component layers. The macroscopic properties show a direct relation with the microscopic crystalline structure. A LGD approach was used to give a semi-quantitative explanation for this behavior taking into account the misfit strains, the electrostatic coupling and the formation of an *a/c*-domain structure. Considering the simplicity of this model the experimental data are well described. The increase of the dielectric constant accompanied by a deterioration of the remnant polarization can be attributed to the changeover from compressive to tensile misfit strain. This was enabled to the observed extent only by the bilayer structure. According to the LGD theory the occurrence of *a*-domains slows down the decrease of the remnant polarization, while the domain walls give a significant contribution to the dielectric constant.

## Acknowledgements

We thank Dr. L. Pintilie for useful hints and fruitful discussions. One of the authors (I.B.M.) wishes to thank the Alexander von Humboldt Foundation for funding his stay in Germany. Work supported by Land Saxony-Anhalt within the Network 'Nanostructured Materials'.